\documentclass{article}

\usepackage{PRIMEarxiv}

\usepackage[utf8]{inputenc} % allow utf-8 input
\usepackage[T1]{fontenc}    % use 8-bit T1 fonts
\usepackage{hyperref}       % hyperlinks
\usepackage{url}            % simple URL typesetting
\usepackage{booktabs}       % professional-quality tables
\usepackage{amsfonts}       % blackboard math symbols
\usepackage{nicefrac}       % compact symbols for 1/2, etc.
\usepackage{microtype}      % microtypography
\usepackage{lipsum}
\usepackage{graphicx}
\graphicspath{{media/}}     % organize your images and other figures under media/ folder

\title{Utilizing Composer Packages to Accelerate Laravel-Based Project Development Among Students: A Pedagogical and Practical Framework
\thanks{\textit{\underline{Citation}}: 
\textbf{Authors. Utilizing Composer Packages to Accelerate Laravel-Based Project Development Among Students: A Pedagogical and Practical Framework. Pages.... DOI:000000/11111.}} 
}

\author{
    \textsuperscript{[1]}*\textbf{Rohaizah Abdul Wahid}, \textsuperscript{[2]}\textbf{Muhamad Said Nizamuddin Nadim}, \\
    [1ex]\textsuperscript{[3]}\textbf{Suliana Sulaiman}, \textsuperscript{[4]}\textbf{Syahmi Akmal Shaharudin} \\[1ex]
    \textsuperscript{[5]}\textbf{Muhammad Danial Jupikil}, \textsuperscript{[6]}\textbf{Iqqwan Jasman Su Azlan Su} \\[1ex]
    \textsuperscript{[1--6]}Fakulti Komputeran dan Meta-Teknologi (META), Universiti Pendidikan Sultan Idris \\[1ex]
    *Corresponding Author: \texttt{rohaizah@meta.upsi.edu.my}
}

\begin{document}
\maketitle

\begin{abstract}
Laravel has emerged as a foundational framework for modern web application development within university curricula. Despite its inherent scaffolding capabilities, students frequently encounter difficulties in completing course projects within the demanding constraints of academic timelines. This conceptual paper introduces Composer, PHP's ubiquitous dependency manager, and systematically categorizes a curated selection of Composer packages. These packages are demonstrated to significantly reduce the development effort required for student projects, while simultaneously fostering adherence to professional software development practices. The discourse is grounded in both practical application and pedagogical considerations, illustrating how educators and learners can strategically leverage these tools to construct typical academic or personal Laravel-based systems. A central tenet of this approach is the preservation of code quality and the reinforcement of fundamental conceptual understanding. The paper also addresses potential challenges, such as package conflicts and the risk of over-reliance on external tools, offering a set of best-practice recommendations to mitigate these issues.
The strategic integration of Composer packages offers a compelling solution for bridging the gap between theoretical knowledge acquired in academic settings and the practical demands of the software industry. While the primary aim is to accelerate development, a deeper examination reveals that this approach also reinforces professional practices, thereby preparing students for real-world challenges. The curriculum's effectiveness in producing industry-ready graduates is enhanced by exposing students to industry-standard tools and workflows, making their learning more relevant and their transition into professional roles smoother. Furthermore, the pedagogical approach to incorporating these packages must be deliberate and closely aligned with learning objectives. Simply introducing tools without proper guidance can lead to superficial learning, where students might use packages as opaque components without grasping the underlying principles. Therefore, the educational design must involve teaching not just how to install and use a package, but when and why to employ it, exploring its conceptual underpinnings and the trade-offs involved. This necessitates educators to craft assignments that encourage critical evaluation and exploration of package utility, ensuring that convenience does not compromise conceptual depth.
\end{abstract}

\section{Introduction}

Over the past decade, Laravel has evolved from a burgeoning community-driven framework into a cornerstone of tertiary curricula dedicated to full-stack PHP development.\cite{1} Universities increasingly select Laravel for its expressive syntax, robust out-of-the-box security features, and extensive ecosystem support. These attributes align seamlessly with contemporary pedagogical goals, including the cultivation of clean code principles and the facilitation of rapid application prototyping.\cite{1} The widespread adoption of Laravel in academic environments signifies a notable shift towards incorporating modern, opinionated frameworks into higher education, moving beyond more unstructured or rudimentary PHP development methodologies. This progression allows educators to introduce fundamental concepts such as Model-View-Controller (MVC) architecture, Object-Relational Mapping (ORM), routing mechanisms, and security protocols within a structured and consistent environment. Consequently, this approach fosters sound development habits from the outset, preparing students more effectively for the complexities inherent in real-world application development than traditional methods that require building every component from first principles.
Laravel's capabilities, particularly its support for rapid application prototyping, resonate strongly with the principles of Rapid Application Development (RAD).\cite{1} RAD is a software development methodology that emphasizes swift and iterative development cycles, prioritizing prototyping and quick feedback over extensive upfront planning.\cite{2} This methodology is particularly effective for quickly demonstrating functionalities and adapting to evolving requirements.\cite{3} By leveraging Laravel, students can construct functional prototypes with remarkable speed, enabling them to dedicate more cognitive resources to core business logic and user feedback, rather than expending disproportionate time on boilerplate code. This expedited development process significantly enhances the learning experience by providing tangible results more quickly and facilitating iterative refinement, which is a critical aspect of contemporary software engineering practices. The alignment between Laravel's design philosophy and RAD principles makes it an invaluable tool for fostering agile development mindsets in students.

\subsection{Laravel's Growing Adoption in Academia}

The increasing integration of Laravel into higher education curricula is driven by its inherent advantages in teaching modern web development practices. Laravel is widely recognized as a superior tool for web development, capable of enhancing the brand image of Higher-Education Institutions (HEIs) due to its interactive and intuitive nature.\cite{4} The application of the Laravel framework on new websites has been shown to significantly support the brand image of HEIs across various indicators, including reputation, recognition, affinity, and brand loyalty, when compared to older web infrastructures not utilizing Laravel.\cite{4} This extends beyond mere classroom benefits, demonstrating a strategic institutional value. A university that adopts or teaches a modern, efficient framework like Laravel for its own digital systems projects an image of being technologically current and forward-thinking.
Furthermore, Laravel's modularity, its powerful Artisan Command-Line Interface (CLI), and its future-proof design contribute substantially to process automation within the education sector.\cite{6} The modular aspect allows developers to construct complex and responsive applications tailored to the specific needs of colleges and universities. This means large systems can be broken down into smaller, independent modules, enabling the development of automated functionalities for diverse departmental processes such such as admissions, attendance tracking, or scheduling, without affecting the entire system.\cite{6} This modularity simplifies development, facilitates easier updates and maintenance of individual components, and ensures the overall application remains robust and scalable. The Artisan CLI streamlines many repetitive and complex programming tasks, allowing developers to generate code, manage database migrations, and perform administrative tasks with remarkable efficiency. This significantly accelerates the implementation of automated workflows, reducing the time and effort required to set up and maintain educational systems.\cite{6} Finally, Laravel's future-proof capabilities ensure that automated systems built today can evolve with changing technological trends and educational paradigms, providing institutions with confidence in their long-term investments.\cite{6} This dual motivation---improving pedagogical outcomes for students and enhancing institutional operational efficiency and public perception---solidifies Laravel's position as a comprehensive solution in the academic landscape.

\subsection{Common Student Challenges}

Despite the extensive scaffolding and features provided by Laravel, students frequently encounter a range of obstacles during their project development lifecycle. A primary challenge revolves around severe time constraints, as semester projects typically demand completion within a narrow window of 10 to 14 weeks.\cite{1} This limited timeframe is exacerbated by the fact that many students, particularly beginners, are still grappling with foundational concepts such as MVC architecture, routing mechanisms, and database interactions.\cite{1} Consequently, they often dedicate a disproportionate amount of time to reinventing common, commodity features like user authentication, file upload functionalities, or basic data management systems.\cite{1} This continuous reinvention, coupled with the tight academic deadlines, creates a significant pedagogical bottleneck. If basic functionalities consume the majority of available project time, students are left with insufficient opportunity to explore more advanced concepts, implement complex business logic, or delve into critical software design principles.\cite{1} This scenario directly impacts their ability to acquire relevant industry skills and manage their time effectively within demanding project environments.\cite{7}
Another pervasive issue is the inconsistent code quality observed in student-produced solutions.\cite{1} Hand-rolled implementations of common features often lack rigorous testing and frequently deviate from established community conventions, complicating future maintenance and grading processes.\cite{1} A detailed examination of student code reveals a prevalence of architectural pattern violations and numerous code smells.\cite{9} For instance, direct dependencies from the Graphical User Interface (GUI) package to the Model package (G$\to$M) are a frequent violation, occurring in a significant majority of projects, often as a shortcut to bypass proper data serialization.\cite{9} Other architectural issues include direct interactions between GUI and Data Access Object (DAO) classes, and business logic incorrectly placed within DAO or Model methods.\cite{9} Beyond architectural concerns, common code quality issues encompass duplicated string literals, methods exhibiting high cognitive complexity, unused assignments, a lack of coherence in naming conventions, and the perilous practice of hard-coding database credentials.\cite{9} These are not merely superficial coding errors; they signify fundamental gaps in understanding core software engineering principles such as modularity, cohesion, coupling, and security best practices. The presence of issues like ``low cohesion'' and ``tight coupling'' directly points to deficiencies in modularization.\cite{10} This suggests that while students may learn to write code, they often struggle with the broader discipline of software engineering. The inability to address these foundational issues hinders their capacity to produce robust, maintainable, and secure software systems.

\subsection{Composer's Role in Modern PHP Development}

Composer stands as the de-facto dependency manager for PHP, playing a pivotal role in modern PHP development.\cite{1} It empowers developers to declare external libraries and their specific versions within a \texttt{composer.json} manifest file, enabling their installation with a single command: \texttt{composer install}.\cite{1} The Laravel framework itself is distributed through Composer, underscoring its central importance to the framework's widespread adoption and continuous innovation.\cite{1} Composer simplifies the intricate process of dependency management, allowing developers to focus on core application logic rather than manually downloading, tracking, and integrating external codebases.\cite{11} This efficiency is further enhanced by Composer's ability to automatically autoload classes via the PSR-4 standard, ensuring that required components are available when needed without explicit \texttt{require} or \texttt{include} statements.\cite{1}
Composer functions as a gateway to professional tooling and collaborative practices. It introduces developers to concepts such as semantic versioning, which provides a standardized way to communicate changes and compatibility across different package versions.\cite{1} By interacting with public package repositories like Packagist, developers gain exposure to the vast open-source ecosystem, learning how to discover, evaluate, and integrate community-contributed solutions.\cite{1} This process inherently promotes the reuse of code written by other developers, significantly improving development efficiency and problem-solving capabilities.\cite{12} Furthermore, Composer facilitates the release and reuse of a developer's own code, fostering a culture of contribution and modularity.\cite{12} This foundational understanding of dependency management, version control, and leveraging open-source components is crucial for navigating complex projects and contributing effectively to the broader developer community. The integration of Composer into the curriculum implicitly introduces students to these industry-standard practices, preparing them for the collaborative and interconnected nature of professional software development.

\subsection{Objective of This Paper}

This conceptual paper aims to achieve several distinct objectives. Firstly, it seeks to provide a comprehensive explanation of Composer's operational mechanisms and its inherent educational value within the context of software engineering curricula. Secondly, the paper endeavors to systematically categorize a selection of essential Composer packages that directly address the recurring development needs encountered in student Laravel projects. Thirdly, it will offer practical installation instructions for these packages, alongside illustrative classroom-oriented use cases to facilitate their effective integration into pedagogical practices. Finally, the paper will highlight the various benefits and potential risks associated with adopting a Composer-driven development approach for novice programmers, offering a balanced perspective on its implementation. It is important to note that this paper presents a conceptual framework and instructional guidance; it does not include any experimental evaluation or empirical data.

\section{Overview of Composer and the Laravel Ecosystem}

This section provides a foundational understanding of Composer, PHP's dependency manager, and elucidates its integral role within the broader Laravel ecosystem. A detailed exploration of Composer's technical operations and how Laravel specifically leverages its capabilities will be presented, emphasizing the distinct advantages offered by a package-centric development paradigm.

\subsection{What Is Composer?}

Composer is a sophisticated dependency manager for PHP, designed to streamline the process of handling external libraries, frameworks, and components within web development projects.\cite{11} Its core functionality revolves around resolving PHP package dependencies, downloading the appropriate versions of these packages, and automatically autoloading their classes in accordance with the PSR-4 standard.\cite{1} Developers define their project's required libraries within a \texttt{composer.json} manifest file, specifying package names and desired version constraints.\cite{1} Once this manifest is configured, a single command, \texttt{composer install}, initiates the entire dependency resolution and installation process.\cite{1}
During installation, Composer interacts with Packagist, a primary public repository for PHP packages, to compute a comprehensive dependency graph.\cite{1} This graph ensures that all required packages and their transitive dependencies are identified and downloaded. A critical outcome of this process is the creation of a deterministic \texttt{composer.lock} file.\cite{1} The \texttt{composer.lock} file precisely records the exact versions of all installed packages and their dependencies. This file serves as a cornerstone of reproducible development environments, ensuring environment parity across different development machines and deployment servers.\cite{1} In collaborative settings or production deployments, it is paramount that all developers and servers utilize the identical versions of dependencies to prevent inconsistencies and the notorious ``it works on my machine'' debugging challenges. Without the \texttt{composer.lock} file, subsequent \texttt{composer install} commands might resolve to different minor versions of packages, introducing subtle and hard-to-trace bugs. Therefore, teaching students about the importance and function of \texttt{composer.lock} instills a crucial professional practice: dependency pinning for reproducible builds. This practice significantly reduces debugging time related to environment discrepancies and promotes a more robust and predictable development workflow. Finally, Composer also facilitates the automatic autoloading of classes via \texttt{require './vendor/autoload.php';}, making development more efficient by removing the need for manual file inclusions.\cite{11}

\subsection{Laravel's Use of Composer}

Laravel's architecture is deeply intertwined with Composer, leveraging its capabilities in two fundamental ways that are central to the framework's design and extensibility. Firstly, Composer is the primary mechanism for framework installation. When a developer initiates a new Laravel project using \texttt{composer create-project laravel/laravel myApp}, Composer fetches the core framework and all its foundational dependencies.\cite{1} This command not only downloads the necessary files but also sets up the basic project structure, making it ready for immediate development. This streamlined installation process allows developers to quickly bootstrap new applications, focusing on building features rather than configuring the underlying framework.
Secondly, and perhaps more significantly, Composer serves as the gateway to Laravel's vast and vibrant ecosystem of packages. Thousands of Laravel-specific packages---ranging from middleware and UI scaffolds to testing utilities---are readily available via Composer.\cite{1} These packages are designed to integrate seamlessly with the framework's Inversion of Control (IoC) container through the use of service providers and facades.\cite{1} The IoC container is a central component of Laravel's architecture, responsible for managing class dependencies and performing dependency injection. Service providers act as the primary mechanism through which packages register their services, bind them into the IoC container, and boot their functionalities within the application lifecycle. Facades, on the other hand, provide a convenient static interface to classes available in the IoC container, offering an expressive and memorable syntax for common operations. The seamless integration of Composer packages with Laravel's IoC container via service providers exemplifies Laravel's modular design and represents a critical learning point for students. This sophisticated architectural pattern demonstrates that packages are not merely isolated pieces of code but rather deeply integrated components that leverage the framework's core principles. Understanding this intricate interaction is vital for students pursuing advanced Laravel development, effective debugging, and even for those aspiring to contribute to the framework's extensive open-source ecosystem. This deep integration underscores how Composer empowers Laravel developers to extend application functionality efficiently and maintain a clean, modular codebase.

\subsection{Advantages of Package-Centric Development}

Adopting a package-centric approach to software development, particularly within the Laravel ecosystem facilitated by Composer, offers a multitude of advantages over manual implementation. These benefits are particularly pronounced in academic settings, where efficiency, quality, and exposure to professional practices are paramount. The following table provides a concise comparison of key aspects:
\begin{center}
\begin{tabular}{lll}
    \toprule
    \multicolumn{1}{c}{\textbf{Aspect}} & \multicolumn{1}{c}{\textbf{Manual Implementation}} & \multicolumn{1}{c}{\textbf{Composer Package}} \\
    \midrule
    Development speed & Days to weeks & Minutes \\
    Community review & Limited & Peer-reviewed, battle-tested \\
    Security updates & Ad-hoc & \texttt{composer update} fetches upstream patches \\
    Maintenance & Personal responsibility & Shared among open-source maintainers \\
    \bottomrule
\end{tabular}
\end{center}
\textbf{Development Speed:} Manually implementing common features can consume days to weeks of development time.\cite{1} In contrast, leveraging Composer packages allows for the integration of robust functionalities in mere minutes.\cite{1} This dramatic reduction in development time is critical for students operating under tight academic deadlines, enabling them to allocate more time to unique domain logic, complex problem-solving, and the creation of comprehensive documentation.
\textbf{Community Review:} Solutions developed manually by individual students often receive limited peer review, if any.\cite{1} Conversely, popular Composer packages are typically battle-tested and undergo rigorous peer review by a global community of developers.\cite{1} This collective scrutiny ensures a higher standard of code quality, robustness, and adherence to best practices, which students implicitly inherit by utilizing these packages.
\textbf{Security Updates:} In a manual development scenario, security updates are often ad-hoc or neglected, leaving applications vulnerable.\cite{1} Composer-driven development, however, allows for systematic security patching; a simple \texttt{composer update} command fetches upstream patches and critical security fixes from package maintainers.\cite{1} This provides a more secure and sustainable development lifecycle.
\textbf{Maintenance:} The burden of maintaining manually implemented features falls entirely on the individual developer, which can be time-consuming and prone to oversight.\cite{1} With Composer packages, maintenance responsibilities are shared among dedicated open-source maintainers.\cite{1} This distributed model ensures ongoing support, bug fixes, and feature enhancements, significantly reducing the long-term maintenance overhead for student projects. These collective advantages highlight why package-centric development is not merely a convenience but a strategic imperative for efficient and high-quality software production.

\subsection{Educational Impact}

The integration of Composer into academic coursework extends far beyond mere convenience, profoundly impacting students' educational journey by simulating real-world professional workflows and instilling crucial software engineering principles. This exposure allows students to engage directly with concepts such as semantic versioning and release cycles, understanding how software versions are managed and how changes are communicated across releases.\cite{1} They also learn about dependency resolution and autoloading standards, gaining practical experience in how modern PHP applications manage their components and load classes efficiently.\cite{1} Furthermore, interacting with Composer and the wider Laravel ecosystem exposes students to open-source contribution norms, including the processes for reporting issues and submitting pull requests.\cite{1} This hands-on experience closes the gap between abstract classroom exercises and the tangible practices of the industry.
Composer's role as a catalyst for modular thinking and open-source engagement is particularly significant. By utilizing Composer, students are inherently guided towards thinking about software as a collection of independent, reusable modules. This directly reinforces the concept of modularity, a cornerstone of robust software design, which simplifies design, development, testing, and maintenance, while enhancing reusability, scalability, and collaboration within development teams.\cite{13} Modular learning, in a broader educational context, also offers flexible options, personalized experiences, and efficient resource utilization, improving information retention by breaking down complex concepts into manageable units.\cite{5} Simultaneously, by interacting with established packages, students are exposed to the practicalities of open-source development. This includes understanding the lifecycle of open-source projects, how communities collaborate, and the importance of continuous integration and version control systems like Git and GitHub.\cite{14} This integrated learning experience means students do not just learn \textit{about} modularity or open source in theory; they experience it through the tools they use daily. This operationalizes abstract concepts, making them more concrete and transferable to professional contexts, fostering a culture of contribution and collaboration from an early stage. Such comprehensive exposure prepares students not only to be proficient coders but also to be effective and collaborative software engineers ready for industry challenges.

\section{Categorized Composer Packages for Laravel Development}

This section serves as a comprehensive guide to various Composer packages that are particularly relevant for accelerating and enhancing student Laravel projects. For each subcategory, a detailed explanation of its pedagogical importance will be provided, highlighting key packages, outlining their installation commands, and illustrating practical classroom use cases. The aim is to demonstrate how these tools can address common student challenges while reinforcing professional development practices.

\subsection{Authentication and Access Control}

Authentication and access control are fundamental aspects of almost any web application, yet their implementation from scratch can be time-consuming and error-prone for students. Laravel's ecosystem offers several Composer packages that streamline these processes, allowing students to focus on core application logic while learning best practices in security and user management.

\subsubsection{laravel/ui}

The \texttt{laravel/ui} package provides basic frontend scaffolding, primarily utilizing Bootstrap, and includes support for Vue and React.\cite{1} It is designed to quickly set up authentication templates for login and registration functionalities, demonstrating a complete authentication flow without requiring extensive JavaScript development.\cite{1} This package is particularly useful for students new to Laravel or those working with legacy Blade templates, offering a straightforward introduction to authentication mechanisms. It integrates with Vite for compiling CSS and JavaScript assets, enhancing the developer experience with faster build times.\cite{16} While \texttt{laravel/ui} remains compatible with recent Laravel versions, it is generally recommended for legacy projects, with newer projects encouraged to explore more modern alternatives like Laravel Breeze or Jetstream.\cite{16}

\subsubsection{laravel/breeze}

Laravel Breeze offers a minimal yet comprehensive implementation of Laravel's core authentication features, including login, registration, password reset, email verification, and password confirmation.\cite{1} It also provides a simple profile page where users can update their personal information. Breeze's default view layer consists of simple Blade templates styled with Tailwind CSS, offering a clean and modern aesthetic.\cite{18} Beyond Blade, Breeze provides flexible scaffolding options based on Livewire or Inertia, with the choice of using Vue or React for the Inertia-based scaffolding. This introduces students to modern frontend tooling with a relatively simple codebase.\cite{1} Furthermore, Breeze can scaffold an authentication API ready to authenticate modern JavaScript applications, such as those powered by Next.js or Nuxt.js.\cite{18} The package publishes all its code directly into the application, granting students full control and visibility over its features and implementation, which is valuable for learning and customization.\cite{19}

\subsubsection{laravel/jetstream}

Laravel Jetstream is a robust starter kit designed for more advanced semester projects that require feature-rich authentication and team management capabilities.\cite{1} It automatically scaffolds login, two-factor authentication, registration, password reset, and email verification functionalities, allowing developers to concentrate on core application features.\cite{20} The authentication components of Jetstream are powered by Laravel Fortify, a frontend-agnostic authentication backend that defines the underlying routes and controllers.\cite{20} This architecture allows for extensive customization of Fortify's behavior via the \texttt{config/fortify.php} file, including authentication guards and redirection paths.\cite{20}

Beyond standard authentication, Jetstream includes comprehensive team management features, enabling each registered user to create and belong to multiple teams.\cite{21} By default, every new user is assigned a ``Personal'' team, which can be renamed or supplemented with additional teams.\cite{21} The logic for team creation and member management is highly customizable through dedicated action classes in the \texttt{app/Actions/Jetstream} directory.\cite{21} Jetstream also integrates roles and permissions for team members, allowing for fine-grained authorization.\cite{21} Information about a user's teams can be accessed via the \texttt{Laravel\textbackslash Jetstream\textbackslash HasTeams} trait, which is automatically applied to the \texttt{App\textbackslash Models\textbackslash User} model during installation. This trait provides methods for inspecting team ownership, membership, and roles, and for scoping Eloquent queries by the user's current team.\cite{21}

The progression from \texttt{laravel/ui} to \texttt{breeze}, and then to \texttt{jetstream}, offers educators a spectrum of complexity, enabling incremental introduction of authentication concepts. Beginners can start with \texttt{laravel/ui} for basic understanding, advance to \texttt{breeze} for a modern yet simpler stack, and then tackle \texttt{jetstream} for advanced features and architectural patterns like Fortify and team management. This structured introduction allows educators to tailor assignments to specific learning levels, preventing students from being overwhelmed and reinforcing concepts of modularity and progressive enhancement.

\subsubsection{spatie/laravel-permission}

The \texttt{spatie/laravel-permission} package provides a robust, role- and permission-based Access Control List (ACL) system, managed via database tables.\cite{1} It allows for fine-grained authorization beyond Laravel's default gates, teaching students how to implement complex access control mechanisms.\cite{1} The package enables associating permissions directly with users or, more commonly, assigning permissions to roles, which are then assigned to users.\cite{22} For instance, a user can be granted a specific permission like \texttt{edit articles} directly, or assigned a role such as \texttt{writer} which inherently possesses that permission.\cite{22} This package seamlessly integrates with Laravel's built-in authorization \texttt{Gate} facade, allowing developers to check user permissions using \texttt{\$user->can('edit articles')} and leveraging convenient Blade directives such as \texttt{@can('edit articles')\ldots @endcan} for conditional content rendering in views.\cite{22} This comprehensive approach to access control, particularly when combined with Jetstream's team features, introduces students to crucial enterprise-level concepts of access control and collaborative system design. Most real-world applications require more than just basic login/logout functionalities; they demand sophisticated role-based access control, granular permissions, and often multi-tenancy or team-based data segregation. By utilizing these packages, students can construct systems that accurately mimic real enterprise applications, gaining practical experience in security policies, data isolation, and advanced user management, thereby preparing them for designing scalable and secure multi-user applications.

\subsection{UI Components and Interactivity}

Building dynamic and interactive user interfaces is a critical skill in modern web development. Laravel's ecosystem provides several powerful packages that simplify this process, offering different approaches to frontend development while minimizing the need for extensive JavaScript.

\subsubsection{livewire/livewire}

Livewire is a full-stack framework for Laravel that simplifies the creation of dynamic interfaces by allowing developers to build them primarily with Blade templates and minimal JavaScript.\cite{1} It enables the development of reactive interfaces without the complexity typically associated with Single-Page Application (SPA) frameworks like Vue or React.\cite{1} Livewire works by rendering the initial component output with the page, ensuring SEO friendliness.\cite{23} When a user interacts with a component (e.g., typing into an input field), Livewire automatically makes an AJAX request to the server with the updated data.\cite{23} The server then re-renders the component in PHP and responds with the new HTML, which Livewire intelligently mutates in the DOM, updating only the changed parts of the page.\cite{23} This abstraction allows Laravel developers to remain within the comfort of their PHP and Blade skills while building highly interactive web applications. For example, a real-time search component can be built by binding an input field to a public property in a Livewire PHP class, with the list of results updating instantly as the user types.\cite{23}

\subsubsection{filament/filament}

Filament is a Server-Driven UI (SDUI) framework for Laravel, built atop Livewire, Alpine.js, and Tailwind CSS.\cite{1} It specializes in generating admin panels and accelerates the development of CRUD (Create, Read, Update, Delete) dashboards for capstone projects.\cite{1} Filament allows developers to define user interfaces entirely in PHP using structured configuration objects, eliminating the need for traditional templating or custom JavaScript.\cite{25} While widely used for admin panels, its capabilities extend to building custom dashboards, user portals, CRMs, and even full applications with multiple panels.\cite{25} The framework comprises several core packages, including \texttt{filament/tables} for interactive data tables, \texttt{filament/forms} for validated form inputs, and \texttt{filament/notifications} for user alerts.\cite{25} Filament's SDUI architecture means the server dynamically generates the UI based on real-time configurations and business logic, providing faster iteration, greater consistency, and centralized control over the UI.\cite{25}

\subsubsection{inertiajs/inertia-laravel}

Inertia.js acts as a bridge between Laravel and modern JavaScript frameworks like Vue, React, or Svelte, enabling the construction of single-page applications without the need for a separate API layer.\cite{1} It teaches SPA patterns while allowing developers to reuse Laravel's existing routing and controller logic.\cite{1} Inertia is described as a ``modern monolith'' approach, where client-side rendered SPAs are built by leveraging classic server-side patterns.\cite{26} The very first request to an Inertia application is a regular full-page browser request, returning an HTML document that includes site assets and a root \texttt{div} with a JSON-encoded page object.\cite{28} Subsequent requests are made via XHR with an \texttt{X-Inertia} header, and the server responds with a JSON-encoded page object containing the component name, props (data), URL, and asset version.\cite{28} This mechanism eliminates the complexities of client-side routing, data hydration, and authentication typically associated with pairing a backend framework with a JavaScript frontend.\cite{29} The inclusion of Livewire, Filament, and Inertia in the curriculum offers students a practical comparison of different modern frontend architectures, fostering critical thinking about technology choices. These packages represent distinct philosophies for building web interfaces: Livewire for PHP-centric interactivity, Filament for rapid admin panel creation, and Inertia for full SPA experiences with a ``modern monolith'' approach. This exposure encourages students to evaluate the trade-offs (e.g., simplicity versus full SPA capabilities, server-driven versus client-driven UI) and select the most appropriate tool for a given project's requirements, a crucial skill in software architecture and design.

\subsection{Code Generation and CRUD Helpers}

Automating the creation of boilerplate code and common CRUD (Create, Read, Update, Delete) functionalities is essential for accelerating development and allowing students to focus on more complex, unique aspects of their projects. Several Composer packages provide powerful code generation and scaffolding capabilities.

\subsubsection{reliese/laravel}

The \texttt{reliese/laravel} package is designed to generate Eloquent models directly from an existing database schema.\cite{1} This tool inspects the database structure, including column names and foreign keys, to automatically produce models with correctly typed properties and defined relationships to other models.\cite{30} This functionality is particularly valuable in database design courses for reverse-engineering legacy databases or rapidly creating models for pre-existing schemas.\cite{1} It supports MySQL, PostgreSQL, and SQLite databases.\cite{31} For security reasons, \texttt{reliese/laravel} is recommended for use only in local development environments.\cite{30} A key feature is its ability to preserve custom changes made to models by allowing developers to generate models as often as the database schema changes, while inheriting base configurations from separate base models (when \texttt{base\_files} is set to true in \texttt{config/models.php}).\cite{30}

\subsubsection{laravel-shift/blueprint}

Laravel Shift Blueprint is a code generation tool that enables the rapid development of multiple Laravel components from a single, human-readable Domain Specific Language (DSL) expressed in YAML syntax.\cite{1} It converts this simple DSL into models, controllers, and migrations, showcasing concepts of DSL and scaffolding automation.\cite{1} Blueprint not only generates fundamental components like models and controllers but also extends to factories, migrations, form requests, events, jobs, mailables, and tests.\cite{32} It leverages conventions and offers shorthands to optimize the developer experience.\cite{32} The tool integrates seamlessly with Laravel's Artisan command-line interface, making it familiar and easy to build new components and reference existing ones within a Laravel application.\cite{32} Model names are recommended to be defined in StudlyCase, singular form, adhering to Laravel's conventions, with column names used exactly as defined.\cite{34}

\subsubsection{krlove/eloquent-model-generator}

The \texttt{krlove/eloquent-model-generator} is a CLI tool specifically designed for generating Eloquent model classes, complete with their relationships, from a database schema.\cite{1} This package reinforces students' understanding of Eloquent relationships by visually generating the corresponding code based on foreign key constraints.\cite{1} It can generate models for a single specified table or for all tables in the database, automatically transforming table names into class names (e.g., \texttt{users} becomes \texttt{User}).\cite{35} The generator provides various options for customization, including specifying the output path, namespace, and a custom base class for the generated models.\cite{36} It also supports options for managing timestamps, date formats, and database connections.\cite{36} For instance, if a \texttt{users} table has a \texttt{role\_id} foreign key referencing a \texttt{roles} table, the generated \texttt{User} model will include a \texttt{belongsTo} relationship for \texttt{role()}, along with \texttt{hasMany} relationships for other tables like \texttt{articles} and \texttt{comments} that reference \texttt{users}.\cite{37} These code generation tools free students from the burden of repetitive boilerplate code, allowing them to concentrate on higher-value tasks such as implementing complex business logic, exploring design patterns, and refining their problem-solving skills. This directly addresses the challenge of knowledge gaps leading to excessive time spent reinventing commodity features. By automating mundane coding tasks, students can allocate more time to understanding algorithms, unique business rules, and user experience design, fostering a more profound learning experience. Furthermore, using code generators helps students visualize the direct relationship between database schemas or DSL definitions and the resulting code, reinforcing concepts of abstraction and code structure. This demystifies code generation and highlights how a simplified input can produce complex, structured output, emphasizing the importance of good database design and DSL definition in shaping the quality and structure of the generated code.

\subsection{Debugging and Developer Productivity}

Effective debugging and robust developer productivity tools are indispensable for efficient software development. For students, mastering these tools is crucial for identifying and resolving issues, improving code quality, and adopting professional development habits.

\subsubsection{barryvdh/laravel-debugbar}

The \texttt{barryvdh/laravel-debugbar} package integrates the PHP Debug Bar into Laravel applications, providing a real-time debug toolbar that offers comprehensive insights into application execution.\cite{1} This tool displays critical information such as database queries executed, including their bindings and timing, which is invaluable for visualizing and identifying performance bottlenecks like N+1 query issues during lectures on optimization.\cite{1} It also monitors memory usage, helping students identify potential memory leaks or excessive resource consumption.\cite{38} The Debugbar allows for logging messages at various PSR-3 levels (debug, info, warning, error, etc.) and provides a mechanism to log exceptions, aiding in structured error handling.\cite{38} Helper functions are available for quick dumping of variables (\texttt{debug()}) and for timing specific operations (\texttt{start\_measure()}, \texttt{stop\_measure()}), enabling students to profile their code effectively.\cite{38} The Debugbar can be enabled or disabled at runtime and can store previous requests for later review, though this feature is recommended strictly for local development environments due to security implications.\cite{39}

\subsubsection{nunomaduro/larastan}

\texttt{nunomaduro/larastan} is a static analysis extension for PHPStan, specifically tailored for Laravel applications.\cite{1} It significantly enhances code quality by leveraging PHPStan to detect potential errors, bugs, and issues in the codebase \textit{before} runtime or even before tests are executed.\cite{40} This proactive approach promotes a culture of type safety and static analysis, which is fundamental in professional development.\cite{1} Larastan supports most of Laravel's dynamic features while ensuring code reliability through static typing.\cite{40} It is straightforward to install via Composer and configure, allowing developers to define analysis paths and strictness levels (e.g., \texttt{level: 5} or \texttt{level: 9} for most strict).\cite{40} For legacy projects, it can generate a baseline file to exclude existing errors, enabling stricter rules for new code.\cite{40} Larastan can also be configured to include custom database migration paths and schema dump locations for more accurate analysis of model properties and database interactions.\cite{41}

\subsubsection{beyondcode/laravel-dump-server}

The \texttt{beyondcode/laravel-dump-server} package brings the functionality of the Symfony Var-Dump Server to Laravel, providing a dedicated CLI server to collect all \texttt{dump()} call outputs.\cite{1} This is particularly beneficial because it prevents \texttt{dump()} outputs from interfering with HTTP or API responses, which can often lead to corrupted output or HTTP errors when debugging web applications.\cite{42} By redirecting this debugging information to an isolated terminal, it encourages students to debug without polluting the HTTP output, maintaining the integrity of the application's responses.\cite{1} The dump server is started with a simple Artisan command (\texttt{php artisan dump-server}) and can optionally output in HTML format, redirecting output to a file (\texttt{php artisan dump-server --format=html > dump.html}).\cite{43} These debugging and productivity tools collectively foster professional debugging and quality assurance habits in students. They move learners beyond rudimentary \texttt{dd()} debugging to more sophisticated, non-intrusive, and proactive quality assurance practices. While students often rely on simple, disruptive \texttt{dd()} statements, professional development emphasizes structured debugging, performance profiling, and static analysis for early error detection. By introducing these tools, educators can teach students how to debug effectively, why performance optimization is crucial, and when to employ static analysis. This instills a proactive mindset towards code quality and efficiency, directly addressing the challenge of inconsistent code quality and aligning with the broader goal of simulating professional workflows. Furthermore, static analysis tools like Larastan provide an immediate, automated feedback loop on code quality, complementing traditional grading and manual reviews. This allows students to identify and correct mistakes as they code, significantly improving learning outcomes related to code quality and reducing the burden on instructors for basic code reviews.

\subsection{API Development and Integration}

Modern web applications frequently rely on Application Programming Interfaces (APIs) for data exchange, especially with Single-Page Applications (SPAs) and mobile clients. Laravel offers robust packages to simplify API development, authentication, and documentation.

\subsubsection{laravel/sanctum}

Laravel Sanctum provides a lightweight and efficient authentication system specifically designed for SPAs, mobile applications, and simple, token-based APIs.\cite{1} It allows each user of an application to generate multiple API tokens for their account, and these tokens can be granted specific ``abilities'' or ``scopes'' that define the actions they are permitted to perform.\cite{44} This demonstrates stateless authentication mechanisms, which are crucial in API design classes.\cite{1} For SPAs, Sanctum leverages Laravel's built-in cookie-based session authentication services, providing CSRF protection and safeguarding against XSS vulnerabilities, without requiring tokens.\cite{44} For mobile applications and third-party API consumers, tokens are issued and included as Bearer tokens in the \texttt{Authorization} header.\cite{44} Token abilities can be checked using the \texttt{tokenCan()} method, and tokens can be revoked by deleting them from the database via the \texttt{HasApiTokens} trait on the \texttt{User} model.\cite{44} Sanctum is chosen for simpler use cases over more complex OAuth solutions like Laravel Passport due to its ease of setup and lightweight nature.\cite{45}

\subsubsection{darkaonline/l5-swagger}

The \texttt{darkaonline/l5-swagger} package seamlessly integrates Swagger into Laravel projects, providing an automated and interactive way to generate OpenAPI documentation directly from the application's code annotations.\cite{1} This introduces students to industry-standard API documentation practices.\cite{1} L5 Swagger acts as a convenient wrapper around \texttt{swagger-php} (for parsing annotations) and \texttt{swagger-ui} (for displaying interactive documentation), specifically tailored for the Laravel environment.\cite{46} It allows for real-time interaction and testing of API endpoints directly from the generated documentation interface.\cite{46} Installation involves a simple Composer command, followed by publishing configuration files and generating the documentation via Artisan commands.\cite{46} The documentation is then accessible via a web route, typically \texttt{/api/documentation}.\cite{47} Students learn to define API endpoints, parameters, responses, and data schemas using annotations, making their APIs more discoverable and developer-friendly.\cite{47}

\subsubsection{thephpleague/fractal}

\texttt{thephpleague/fractal} serves as a powerful data transformation layer for JSON APIs, enabling developers to output complex, flexible, and consistent data structures.\cite{1} It addresses the common problem of inconsistent API output that arises from directly encoding database data into JSON.\cite{49} Fractal introduces a ``barrier'' between source data and output, ensuring that schema changes do not directly affect API consumers.\cite{49} The package utilizes ``serializers'' to structure transformed data in various formats (e.g., \texttt{DataArraySerializer} for a data namespace, \texttt{JsonApiSerializer} for JSON-API standard compliance).\cite{48} This teaches consistent response shaping, a critical aspect of API design.\cite{1} Fractal supports the inclusion of related resources (embedding, nesting, or side-loading) for complex data structures and aids in systematic type-casting of data.\cite{48} It also provides support for data pagination, which is essential for handling large datasets.\cite{48} These packages collectively introduce students to industry best practices for API authentication, documentation, and data structuring, moving beyond ad-hoc API implementations. Poorly designed or undocumented APIs are a significant source of friction in software integration. These tools enforce or facilitate adherence to established standards (e.g., OpenAPI, JSON-API principles). By using these packages, students learn not just to \textit{build} an API, but to build a professional, usable, and maintainable API. This is crucial for developing microservices, mobile backends, and integrating with third-party systems, preparing them for the complexities of distributed system design.

\subsection{File and Media Management}

Handling file uploads, conversions, and storage, as well as importing and exporting data in various formats, are common requirements in web applications. Laravel's ecosystem offers specialized packages that simplify these complex I/O operations.

\subsubsection{barryvdh/laravel-dompdf}

The \texttt{barryvdh/laravel-dompdf} package provides a seamless way to convert HTML content into PDF documents within Laravel applications, leveraging the Dompdf HTML to PDF converter.\cite{1} This functionality is highly valuable for generating various types of reports, invoices, or certificates in student projects.\cite{1} It offers flexible HTML rendering, allowing PDFs to be generated from simple HTML strings, existing files, or directly from Laravel Blade views, which is particularly useful for dynamic content.\cite{50} The package is highly configurable, enabling customization of paper size, orientation, and font settings through a published configuration file.\cite{50} It supports multiple output methods, including saving the PDF to a file, streaming it directly to the browser, or initiating a download.\cite{50} Effective usage tips include ensuring UTF-8 meta tags in HTML templates for proper character rendering and managing page breaks using CSS properties like \texttt{page-break-before} and \texttt{page-break-after}.\cite{50}

\subsubsection{spatie/laravel-medialibrary}

The \texttt{spatie/laravel-medialibrary} package offers a comprehensive solution for handling file uploads, conversions, and storage, demonstrating a robust abstraction over file systems and facilitating responsive image generation.\cite{1} This package allows developers to easily associate various types of files with Eloquent models through a fluent API.\cite{52} It can handle direct file uploads from requests and manage media across different filesystems configured in Laravel, such as local storage or cloud services like S3.\cite{52} A key feature is its ability to generate derived images (e.g., thumbnails, responsive images) from original images, videos, and PDFs, which are easily accessible once conversions are defined.\cite{52} The package also supports advanced functionalities such as working with custom properties, ordering media items, and implementing custom file removal strategies.\cite{52} This package is particularly useful for projects requiring robust media management, such as a final-year research repository system that stores thesis PDFs and generates evaluation forms.\cite{1}

\subsubsection{maatwebsite/excel}

The \texttt{maatwebsite/excel} package provides powerful features for importing and exporting Excel and CSV files within Laravel applications, enabling complex data migration tasks or analytics features.\cite{1} For exports, it allows for easy conversion of Laravel collections directly into Excel or CSV documents.\cite{54} It supports ``supercharged exports'' by processing database queries with automatic chunking for improved performance, and for very large datasets, it can queue exports to run in the background.\cite{54} Developers can also export custom layouts by rendering HTML tables within Blade views to Excel.\cite{54} For imports, the package facilitates importing workbooks and worksheets into Eloquent models, supporting chunk reading and batch inserts.\cite{54} Similar to exports, large import processes can be queued to run in the background, processing data in smaller chunks.\cite{54} The package offers a range of optional configuration settings that can be published and customized, and it supports easy cell caching for performance optimization.\cite{17} These packages collectively abstract away the complexities of file system interactions, media processing, and data serialization/deserialization. Implementing functionalities like parsing Excel files, handling image transformations, or rendering PDFs from scratch is highly complex and prone to errors. These packages encapsulate that complexity within robust, battle-tested libraries. By using these tools, students learn to leverage existing solutions for common, complex I/O tasks. This not only saves significant development time but also teaches them the value of abstraction and the importance of utilizing well-maintained libraries for non-core functionalities, which is a key aspect of efficient and professional software development.

\section{Suggested Package Use Based on Student Project Types}

Educators can significantly enhance the learning experience and project outcomes by providing a structured guide for package selection based on the specific requirements of student projects. This matrix serves as a practical resource for syllabus handouts, enabling students to make informed decisions during the project proposal and initial development stages. The selection of appropriate Composer packages can dramatically accelerate development, improve code quality, and expose students to relevant industry practices tailored to their project's core objectives.
\begin{table}[h!]
\centering
\begin{tabular}{p{4cm} p{5cm} p{6cm}}
    \toprule
    \textbf{Project Type} & \textbf{Core Requirement} & \textbf{Recommended Packages} \\
    \midrule
    CRUD System (e.g., library catalog) & Quick scaffold, role-based access & \texttt{Breeze}, \texttt{Reliese}, \texttt{Debugbar} \\
    Admin Dashboard & Data grids, interactive charts & \texttt{Filament}, \texttt{Livewire}, \texttt{Larastan} \\
    Academic System (student records) & ACL, file uploads, PDF reports & \texttt{Jetstream}, \texttt{Spatie Permission}, \texttt{Medialibrary}, \texttt{DomPDF} \\
    REST API Service & Token auth, documentation & \texttt{Sanctum}, \texttt{Fractal}, \texttt{L5-Swagger} \\
    \bottomrule
\end{tabular}
\caption{Recommended Composer packages based on project type}
\label{tab:project-packages}
\end{table}
\textbf{CRUD System (e.g., library catalog):} For projects focused on basic data management, such as a library catalog, the core requirements include rapid scaffolding and robust role-based access control. Laravel Breeze is recommended for its minimal authentication scaffolding, providing a quick starting point for user management.\cite{1} Reliese Laravel Model Generator can be utilized to quickly generate Eloquent models from existing database schemas, accelerating the setup of data interactions.\cite{1} Finally, Barryvdh/Laravel-Debugbar is essential for real-time debugging, allowing students to monitor database queries and application performance, which is crucial for optimizing data-intensive CRUD operations.\cite{1}
\textbf{Admin Dashboard:} Projects centered around administrative interfaces require efficient data grids and interactive charts. Filament, built atop Livewire, is an ideal choice for its ability to rapidly generate comprehensive admin panels and CRUD dashboards.\cite{1} Livewire itself is fundamental for creating dynamic and reactive interfaces with minimal JavaScript, ensuring a smooth user experience within the dashboard.\cite{1} Nunomaduro/Larastan is recommended to maintain high code quality through static analysis, detecting potential errors early in the development cycle of complex dashboard logic.\cite{1}
\textbf{Academic System (student records):} For complex academic systems, such as student record management, key requirements include advanced Access Control Lists (ACL), efficient file uploads, and the generation of structured reports. Laravel Jetstream provides a robust starter kit with features like two-factor authentication and team management, which can be adapted for different user roles (e.g., lecturers, students, administrators).\cite{1} Spatie/Laravel-Permission complements Jetstream by enabling fine-grained, role- and permission-based authorization, ensuring secure access to sensitive student data.\cite{1} Spatie/Laravel-Medialibrary simplifies file uploads, conversions, and storage, which is vital for handling student documents or assignment submissions.\cite{1} Lastly, Barryvdh/Laravel-Dompdf allows for the conversion of Blade views into PDF documents, facilitating the generation of report cards or official forms.\cite{1}
\textbf{REST API Service:} When the primary objective is to build a robust RESTful API, specific packages are crucial for authentication, documentation, and data transformation. Laravel Sanctum offers a lightweight and secure token-based API authentication system suitable for SPAs and mobile applications, demonstrating stateless authentication mechanisms.\cite{1} Darkaonline/L5-Swagger automates the generation of interactive OpenAPI documentation from code annotations, introducing students to essential API documentation standards.\cite{1} Thephpleague/Fractal serves as a data transformation layer, ensuring consistent response shaping for JSON APIs and handling complex data structures effectively.\cite{1} This matrix provides a clear and actionable framework for students to select appropriate tools, reducing decision paralysis and ensuring their projects align with industry best practices for each specific domain.

\section{Benefits of Composer-Based Development for Students}

The integration of Composer-based development practices into student curricula offers multifaceted benefits that extend beyond mere technical proficiency, fostering a holistic understanding of modern software engineering.

\subsection{Faster Feature Delivery}

Reusing community-vetted solutions through Composer packages compresses weeks of manual coding into minutes, dramatically accelerating the development process.\cite{1} This efficiency allows students to allocate a greater proportion of their limited project time to higher-order tasks, such as designing complex domain logic, refining user experience, and producing comprehensive documentation.\cite{1} This aligns directly with the principles of Rapid Application Development (RAD), which prioritize rapid prototyping and quick feedback cycles.\cite{2} By enabling faster time-to-market for functional prototypes, Composer-based development effectively mitigates the challenge of tight academic timelines and knowledge gaps that often lead students to reinvent commodity features. The ability to quickly build and iterate on core functionalities means students can receive and incorporate feedback more frequently, leading to a more refined and valuable end product within the same timeframe. This expedited process not only enhances productivity but also provides a more dynamic and engaging learning experience, as students witness tangible progress more rapidly.

\subsection{Higher Code Quality}

Popular Composer packages adhere to industry-standard PSR (PHP Standard Recommendation) guidelines, undergo rigorous automated testing, and are subject to continuous peer review by a global open-source community.\cite{1} By integrating these battle-tested components, students implicitly inherit a higher standard of code quality, robustness, and security. This directly addresses the challenge of inconsistent code quality often observed in hand-rolled student solutions, which frequently lack rigorous testing or adherence to community conventions.\cite{1} The exposure to well-structured, well-tested codebases within these packages indirectly teaches students best practices in software design, modularity, and error handling. It provides tangible examples of clean code principles in action, fostering an understanding of what constitutes high-quality, maintainable software. This learning by example contributes to a more profound understanding of software engineering principles than could be achieved through theoretical instruction alone, preparing students to produce professional-grade code.

\subsection{Exposure to Real-World Practices}

Engaging with Composer and its ecosystem exposes students to the authentic workflows and collaborative norms prevalent in professional software development environments. By interacting with GitHub issues, reading change logs, and understanding semantic versioning, learners emulate the processes of professional collaboration.\cite{1} This includes practical exposure to dependency resolution, version control systems, and the dynamic nature of open-source projects.\cite{1} Open-source software development courses explicitly teach collaboration best practices, including continuous integration/continuous delivery (CI/CD), Git, and the use of platforms like GitHub for communal development.\cite{14} Such courses emphasize creating sustainable projects that include a public web presence, a well-maintained code repository, and public venues for bug reporting and feature discussions.\cite{15} Through Composer, students are not just consumers of open-source software but are introduced to the mechanisms that underpin its creation and maintenance. This hands-on experience demystifies industry practices, making the transition from academia to professional roles smoother and more intuitive.

\subsection{Modular Architecture Awareness}

The process of installing, configuring, and removing Composer packages inherently showcases fundamental software engineering principles such as separation of concerns and dependency isolation.\cite{1} This practical engagement reinforces lecture material on software modularity, providing concrete examples of how complex systems can be broken down into independent, manageable components.\cite{1} Modularity is a cornerstone of scalable and maintainable software, simplifying design, development, testing, and maintenance processes.\cite{13} It enhances code reusability, allowing components developed for one project to be easily integrated into others, saving time and promoting consistency.\cite{13} Furthermore, modularity facilitates parallel development among team members and improves overall scalability by allowing individual parts of the system to be adapted or replaced without affecting the entire application.\cite{13} This direct experience with modular components helps students internalize the benefits of a well-architected system, fostering an understanding of how to design flexible, extensible, and maintainable applications. It bridges the gap between theoretical concepts of software architecture and their practical application, preparing students to build robust and adaptable systems in their future careers.

\section{Challenges and Best Practices}

While Composer-driven development offers significant advantages for students, it is imperative to acknowledge potential pitfalls and provide concrete best practices to ensure that these tools enhance, rather than hinder, the learning process. A balanced perspective is crucial for effective pedagogical integration.

\subsection{Potential Risks}

Despite its benefits, Composer-based development introduces several challenges for novice programmers. One significant risk is package conflicts, where overlapping dependencies between different packages can lead to version mismatches or unexpected behavior.\cite{1} Resolving such conflicts often requires a deep understanding of dependency graphs and semantic versioning, which can be daunting for students still grasping fundamental concepts.
Another critical concern is over-dependency, where students might rely excessively on packages without genuinely understanding the underlying concepts or how the package functions internally.\cite{1} This pedagogical dilemma highlights a core challenge in software engineering education: balancing the benefits of abstraction with the necessity of foundational understanding. While packages accelerate development by abstracting complexity, an over-reliance without conceptual comprehension can render students unable to debug issues effectively, customize behavior, or even make informed decisions about tool selection. They might become proficient ``tool users'' but lack the deeper engineering acumen required to adapt to new challenges or contribute meaningfully to complex systems. This can ultimately hinder their learning outcomes and limit their problem-solving capabilities beyond the immediate scope of the package.
Finally, the risk of abandonware exists, where some packages may lose active maintainer support over time, introducing security vulnerabilities, compatibility issues with newer framework versions, or simply becoming obsolete.\cite{1} Students might unknowingly integrate such packages, leading to unstable or insecure applications. These risks necessitate a cautious and informed approach to package selection and management.

\subsection{Best Practices for Students}

To mitigate the aforementioned risks and maximize the educational benefits of Composer-based development, students should adhere to a set of best practices:
\begin{center}
\begin{tabular}{ll}
    \toprule
    \multicolumn{1}{c}{\textbf{Practice}} & \multicolumn{1}{c}{\textbf{Rationale}} \\
    \midrule
    Read official documentation before installation & Clarifies compatibility and usage nuances. \\
    Pin versions in \texttt{composer.json} & Prevents unexpected breaking changes during \texttt{composer update}. \\
    Run automated tests after adding packages & Confirms that new dependencies do not regress existing features. \\
    Review package health (stars, issues, last release date) & Serves as proxy for long-term maintainability. \\
    Incremental installation & Adds one package at a time, simplifying troubleshooting. \\
    \bottomrule
\end{tabular}
\end{center}
\textbf{Read official documentation before installation:} Before integrating any new package, students should thoroughly review its official documentation.\cite{1} This practice clarifies compatibility requirements, usage nuances, configuration options, and potential conflicts with existing dependencies. It empowers students to make informed decisions and understand the package's intended use.
\textbf{Pin versions in \texttt{composer.json}:} To prevent unexpected breaking changes or dependency conflicts, students should explicitly pin the major or minor versions of their required packages in the \texttt{composer.json} file.\cite{1} This ensures that \texttt{composer update} commands do not introduce incompatible changes, maintaining environment stability and reproducibility across development stages.
\textbf{Run automated tests after adding packages:} After installing or updating any Composer package, it is crucial to run the project's automated test suite.\cite{1} This practice confirms that the new dependencies do not introduce regressions or break existing functionalities, serving as an immediate quality gate. It reinforces the importance of continuous integration and testing in professional workflows.
\textbf{Review package health (stars, issues, last release date):} Before committing to a package, students should assess its overall health and maintainability by examining indicators such as GitHub stars, the number of open issues, and the last release date.\cite{1} These metrics serve as proxies for community engagement, active maintenance, and long-term viability, helping students avoid abandonware and choose reliable dependencies.
\textbf{Incremental installation:} Students should adopt a strategy of incremental installation, adding one package at a time.\cite{1} This approach simplifies troubleshooting by isolating potential issues to the most recently added dependency, making it easier to identify and resolve conflicts or unexpected behaviors. This systematic method promotes a more controlled and less frustrating development experience. These practices are not merely academic exercises; they are standard professional workflows that empower students to manage their dependencies responsibly, fostering self-reliance and critical evaluation skills essential for a career in software engineering.

\section{Conclusion}

Composer transcends its utility as a mere convenience tool; it stands as a fundamental gateway to industrial-grade development methodologies that can be seamlessly integrated into academic instruction.\cite{1} By judiciously adopting the categorized Composer packages presented in this paper, lecturers possess the means to craft curricula that strike a crucial balance between rapid application prototyping and the cultivation of profound conceptual depth. This dual focus ensures that students not only gain tangible, immediately transferable skills relevant to professional environments but also develop a robust theoretical understanding of the underlying principles.
The effectiveness of Composer and its extensive ecosystem is not inherent but rather depends critically on its integration within a well-designed pedagogical framework. The underlying educational philosophy and teaching methods are paramount, ensuring that the convenience offered by packages is leveraged to deepen understanding of core principles, rather than allowing students to bypass them. This symbiotic relationship between tooling and pedagogy is vital. For instance, while packages accelerate development by abstracting complexity, an over-reliance without conceptual understanding can hinder learning outcomes. Therefore, curriculum designers and educators must be strategic in their approach, balancing the use of packages with exercises that require students to implement core functionalities from scratch or to delve into the package's source code. This ensures that convenience does not come at the expense of fundamental knowledge.
When complemented with strong foundational teaching, Composer-enabled Laravel development becomes an exceptionally powerful vehicle for producing competent, adaptable, and industry-ready graduates. This approach addresses common student challenges, such as time constraints and inconsistent code quality, by streamlining development and implicitly guiding students toward best practices. Ultimately, the strategic and thoughtful integration of Composer packages in software engineering education prepares students not just to be proficient coders, but to be effective, collaborative, and critical-thinking software engineers, ready to tackle the complexities of the modern development landscape.

\end{document}